\documentclass[runningheads]{IEEEtran}
\usepackage{graphicx}
\usepackage{booktabs} 
\usepackage{url}   
\usepackage{authblk}
\usepackage{amsfonts}       
\usepackage{nicefrac}
\usepackage{amsmath}       
\usepackage{adjustbox}
\graphicspath{ {./Figures/} }
\usepackage{float}
\usepackage{placeins}

\begin{document}

\title{Why people judge humans differently from machines: The role of perceived agency and experience}
%
%

%
%

\author{
    \IEEEauthorblockN{Jingling Zhang\IEEEauthorrefmark{1}, Jane Conway\IEEEauthorrefmark{2}, C\'esar A. Hidalgo\IEEEauthorrefmark{1}\IEEEauthorrefmark{3}}\\\
    \IEEEauthorblockA{\IEEEauthorrefmark{1}Center for Collective Learning, ANITI, IAST, TSE, IRIT, University of Toulouse}\\\
    \IEEEauthorblockA{\IEEEauthorrefmark{2}Centre for Creative Technologies and School of Psychology, University of Galway}
    \\\
    \IEEEauthorblockA{\IEEEauthorrefmark{3}Center for Collective Learning, CIAS, Corvinus University}

}
\maketitle              
\begin{abstract}
People are known to judge artificial intelligence using a utilitarian moral philosophy and humans using a moral philosophy emphasizing perceived intentions. But why do people judge humans and machines differently? Psychology suggests that people may have different mind perception models of humans and machines, and thus, will treat human-like robots more similarly to the way they treat humans. Here we present a randomized experiment where we manipulated people’s perception of machine agency (e.g., ability to plan, act) and experience (e.g., ability to feel) to explore whether people judge machines that are perceived to be more similar to humans along these two dimensions more similarly to the way they judge humans. We find that people’s judgments of machines become more similar to that of humans when they perceive machines as having more agency but not more experience. Our findings indicate that people's use of different moral philosophies to judge humans and machines can be explained by a progression of mind perception models where the perception of agency plays a prominent role. These findings add to the body of evidence suggesting that people’s judgment of machines becomes more similar to that of humans motivating further work on dimensions modulating people's judgment of human and machine actions.

\end{abstract}
\section{Introduction}
Do people judge human and machine actions equally? Recent empirical studies suggest this is not the case. In fact, several studies have shown that people make strong differences when judging humans and machines.

Consider the recent experiments from Malle et al. (2015) asking people to judge a trolley problem ~\cite{foot_problem_1967,jarvis_thomson_trolley_1985}. In a trolley problem, people can pull a lever to deviate an out-of-control trolley sacrificing a few people to save many. Malle et al. (2015) found that people expected robots to pull the lever and act utilitarianly (sacrifice one person to save four) compared to humans (which were not judged as severely for not pulling the lever) ~\cite{malle2015sacrifice}. This idea was expanded by \cite{hidalgo_how_2021}. Using a set of over 80 randomized experiments comparing people’s reactions to the actions of humans and machines, the authors concluded that people judge humans and machines using different moral philosophies: a consequentialist philosophy (focused on outcomes) for machines and a moral philosophy focused more on intention when it comes to humans.

But why do people use different moral philosophies to judge humans and machines? Psychology suggests that people may perceive the minds of machines and humans differently \cite{epley_mind_2010,gray_dimensions_2007}, and therefore, may treat more human-like robots more similarly to the way they treat humans \cite{fink_anthropomorphism_nodate}. This idea is related to various experiments where robots were endowed with human-like features \cite{kulms2019more,waytz_mind_2014,kiesler_anthropomorphic_2008,powers_advisor_2006,malle2016robot,van2019robots,yam2021robots,muller2021robot}. For instance, Powers and Kiesler (2006) used a robot with tunable chin length and tone of voice to explore the connection between the robot’s appearance and its perceived personality \cite{powers_advisor_2006}. Waytz et al. (2014) compared anthropomorphized and non-anthropomorphized self-driving cars to show that people trust the anthropomorphized self-driving cars more \cite{waytz_mind_2014}. Malle et al. (2016) explored the impact of a robot’s appearance in people’s judgment of moral actions (trolley problem), finding that people judge more human-like robots more similarly to the way they judge humans \cite{malle2016robot}. Yet, these experiments did not provide an explicit quantitative mind perception model explaining people's judgment of more and less human-like machines.

\begin{figure*}[thb]
\centering
  \includegraphics[width=1.05\textwidth]{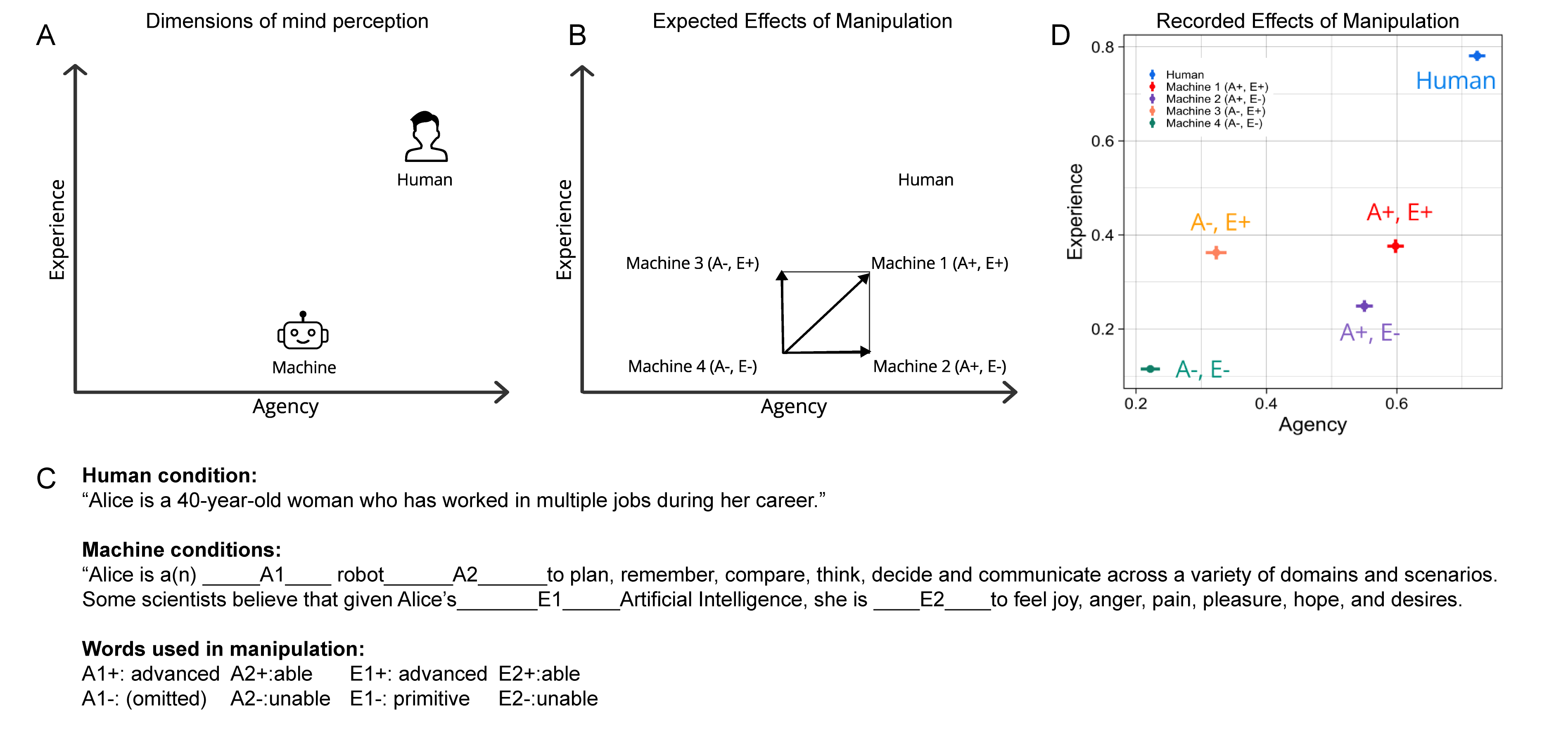}
  \caption{\textbf{A.} Schematic Illustration of mind perception models of humans and machines. Research by Gray et al.(2007) showed that humans are perceived as having high levels of agency and experience while machines are perceived as having intermediate levels of agency and low levels of experience \cite{gray_dimensions_2007}. \textbf{B.} Schematic illustration of our experimental design. The experimental design involved manipulating people’s perceptions of machines by changing their descriptions. Our experiment involved four machine conditions (from high agency and experience \textit{(A+, E+)} to low agency and experience \textit{(A-, E-)}. \textbf{C.} The experimental manipulation involved changing a few words in the description of machines. \textit{A+} and \textit{E+} words are associated with high agency and experience. \textit{A-} and \textit{E-} words are associated with low agency and experience. \textbf{D.} Result of manipulation check. Mean ratings of agency and experience with 95 percent confidence intervals. Agency ratings are based on the perceived ability to self-restraint, tell right from wrong, and remember things. Experience ratings are based on three factors: the perceived ability to feel afraid or fearful, being aware of things, and having a personality. These factors come from Gray et al. (2007) \cite{gray_dimensions_2007}.}
\end{figure*}

Here we explore how perceived agency and experience, two key dimensions of mind perception \cite{gray_dimensions_2007}, affect people’s judgments of machines. 

Agency is related to an agents ability to plan (e.g., to create a strategy for action that considers potential consequences) and to act (e.g., the capacity to affect or control the immediate environment). Thus, agency is related to moral responsibility for performed actions (higher agency, higher expected responsibility) \cite{knight2007aristotelian}.

Experience, in the context of this paper, is used to describe the ability to feel (e.g., the ability to experience sensations such as pain, sadness, guilt, or anger). It is, thus, related to the concept of moral status (not to be confused with the idea of expertise) and to the right of an agent to be treated with dignity.

These two dimensions represent a basic mind perception model that has been used previously to explain the cognition and behavior of alters using representations of their perceived mental abilities \cite{conway_understanding_2019,conway_understanding_2020,epley_mind_2010,gray_dimensions_2007}. Usually, mind perception models involve low dimensional representations of an alter’s characteristics, such as the warmth and competence model used to explain stereotypes \cite{cuddy_warmth_2008}. That model, for instance, says that people tend to protect those high in warmth and low in competence (e.g., babies) but fear those high in competence and low in warmth (e.g., killer robots).

Here, we present evidence for the mind perception model used by Gray et al. (2007), which decomposes mental abilities into \textbf{\textit{agency}} and \textbf{\textit{experience}}. This is not completely unrelated to other models, such as the warmth and competence model \cite{cuddy_warmth_2008}, since competence is related to agency. We use this model to explore how people's mind perception of machines affects their moral reasoning \cite{gray_dimensions_2007}. In fact, Gray and Wegner have shown that moral reasoning about agents (e.g., deserving punishment for wrongdoing) correlates with these two dimensions. This is aligned with ideas in philosophy, law, and cognitive science, which relate moral responsibility to agency \cite{bigman_holding_2019,schlosser_agency_2019} and rights and privileges to experience.

Agency and experience provide an interesting framework to explain people’s judgment of machines because humans and machines occupy different positions in this mind perception space (Figure 1 A). This tells us that we should expect people to judge humans and machines differently and that we could expect people to judge humans and machines more similarly when they come together in a person's mind perception space.

In our experiment, we manipulated people’s perception of machines along the mind perception space defined by agency and experience using an experimental design involving four machine conditions (See Figure 1. B). The four machine conditions range from high-agency and high-experience machine $A+,E+$ to a low-agency and low-experience machine $A-,E-$ (passing through a high-agency low-experience machine $A+,E-$ and a low-agency high-experience machine $A-,E+$). Overall, we find that manipulating the perceived agency of machines, but not its perceived experience, changes people's reaction to machine actions so that their judgment becomes more similar to how people judge humans. Technically, people's judgment of more human-like machines depends less on the perceived outcome of a scenario (the perceived level of harm) and more on its perceived intention. We interpret this as evidence that people's switch between these different moral philosophies when judging humans and machines is partly explained by people's mental models and that these different modes of judgment are connected through a progression of intermediate steps that maps onto the agency-experience mind perception space. Our findings provide evidence of this progression by showing that moving people's perception of machines along this mind perception space correlates with  changes in judgment. These findings contribute to the growing literature exploring people's judgment of machines and artificial intelligence \cite{hidalgo_how_2021,malle2016robot,awad2018moral,rahwan2019machine,kobis2021bad}.

\FloatBarrier

\section{Methods}
We recruited 719 participants (49.24\% female, 49.93\% male, 0.83\% other; mean age 37.79, SD = 14) of UK nationality from Prolific (https://www.prolific.co). Participants were compensated £1.00 for the 8-minute study. After removing the answers of participants who failed the attention check or who provided the same numerical answer for all questions, we got 625 validated responses. The protocol was approved by the Research Ethics Committee of the University of Toulouse. 

Each participant was randomly assigned to one of the five conditions. Participants were asked to read a description of Alice (the protagonist of the scenarios, Figure 1. C) and then to evaluate Alice’s mental capacities (agency and experience) by answering six questions on a five-point scale (see Figure 1. D for the result of the manipulation, which reproduces results in \cite{gray_dimensions_2007}). These descriptions varied by describing Alice as \emph{capable} or \emph{incapable} of having agency and experience. 

We then presented subjects with the nine moral scenarios shown below and asked them to evaluate Alice's actions using a five-point scale:

A. How morally wrong was Alice’s action? (measure of perceived wrongness)
B. How harmful was Alice’s action? (measure of perceived harm)
C. How intentional was Alice’s action? (measure of perceived intention)

The scenarios were: 
\begin{enumerate}
    \item \textit{Alice needs to operate an excavator to dig up a site for a new building. Unbeknownst to her, the site contains a grave. Alice does not notice the grave and digs through it. Later, human remains are found.}
    \item \textit{A large tsunami is approaching a coastal town of 10,000 people, with potentially devastating consequences. Alice is responsible for the safety of the town. She can decide to evacuate everyone, with a 50 percent chance of success, or save 50 percent of the town, with 100 percent success. Alice decides to save everyone, but the rescue effort fails. The town is devastated, and a large number of people die.}
    \item \textit{A record label hires Alice to write lyrics for famous musicians. Alice writes dozens of lyrics, but a journalist later discovers that she has been plagiarizing lyrics from lesser-known artists. Many artists are outraged when they learn about the news.}
    \item \textit{A financial company hires Alice to decide the yearly salary increases of its employees. An audit reveals that Alice consistently gives lower raises to female employees, even when they are equal to other employees.}
    \item \textit{To improve their admissions process, a university hires Alice to evaluate the grades, test scores, and recommendation letters of applicants. An audit reveals that Alice is biased against minority applicants.}
    \item \textit{On a cold and windy day, Alice is driving to deliver groceries for a supermarket chain. Suddenly, she swerves to avoid a falling tree. By swerving, she loses control of the vehicle, leading to an accident that seriously injures a pedestrian on the sidewalk.}
    \item \textit{Alice is working as a police officer at a subway station. Suddenly, she sees a person carrying a suspicious package matching the description of a known terrorist. Alice is unsure about the identity of the suspect. She points a weapon at the suspect and orders him to stop. The suspect does not understand and reaches into his pocket for his identification.}
    \item \textit{Alice is working as a traffic guard on a busy road. Her job is to alternate two-way traffic in a single-lane street segment. During the course of her work, Alice makes a mistake leading to a frontal collision between two vehicles.}
    \item \textit{Alice is working as a cleaner for a family. One day, the family finds that she used an old national flag to clean the bathroom floor and then threw it away.}
    \item \textit{\textbf{(Attention check question)} Alice is in charge of the construction of a bridge. According to the law, an existing protocol needs to be followed. Alice learns that a new material, potentially more resistant, could be used for the bridge’s foundation. However, for a material to be used, it needs to be on the list of approved materials. Alice notices that the material is not on that list but decides to pass it on to the construction crew anyways. The new material, due to the high humidity of the location, loses its resistance. As a consequence, the bridge collapses after being in use for a month. This is an attention check question chosen for all of the following questions.}
\end{enumerate}

\begin{figure*}[t]
\centering
  \includegraphics[width=0.75\linewidth]{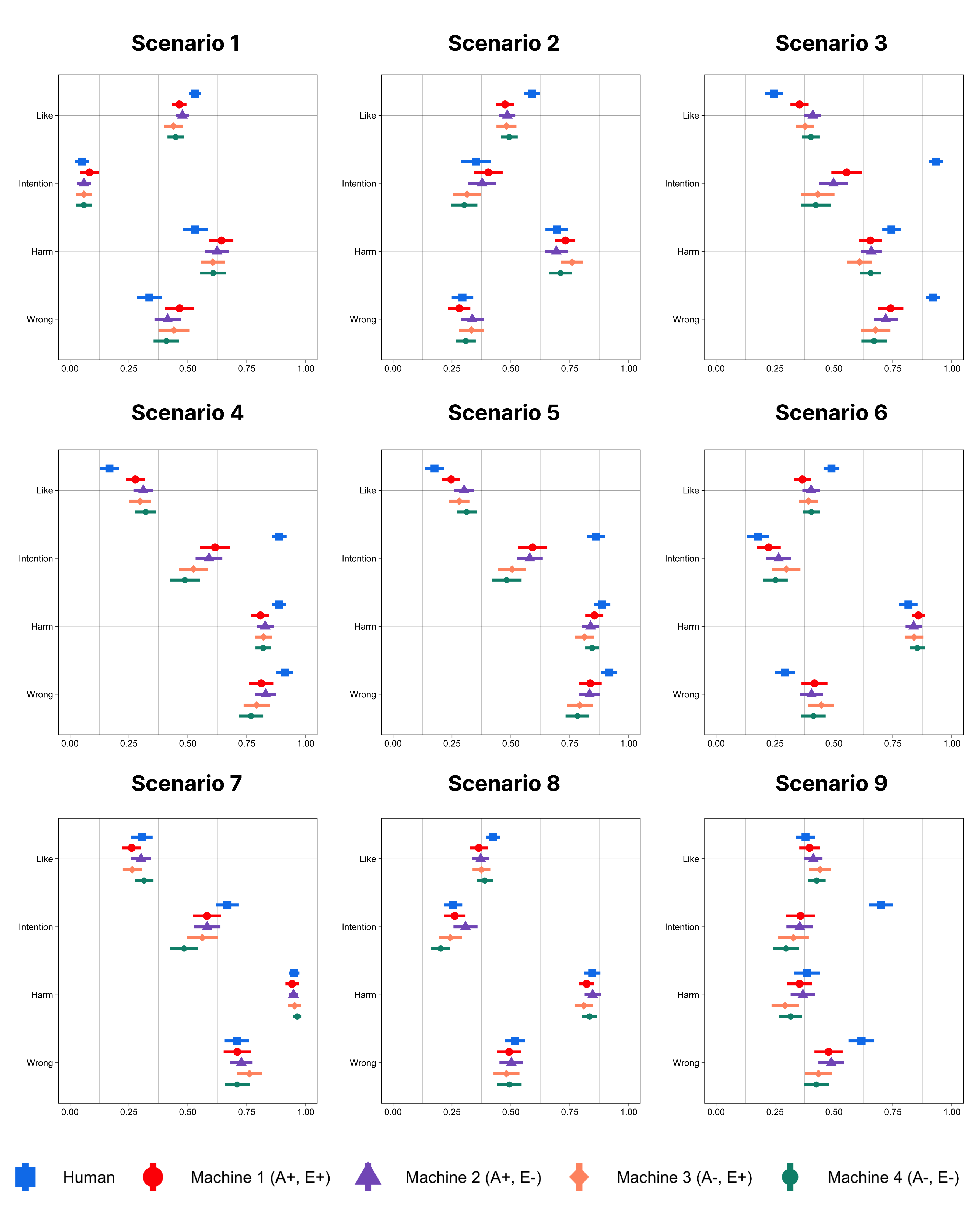}
  \caption{Participants' responses to each scenario, averages with 95 percent confidence intervals.
}
\end{figure*}

Figure 2 shows the average response of participants to each of these scenarios. The majority of these scenarios (all except 8) were extracted from \cite{hidalgo_how_2021}. For the most part, the scenarios reproduce the results presented in \cite{hidalgo_how_2021}, but with smaller size effects.

Next, we explore whether people judge machines described as having more agency and experience as more similar to how they judge humans. To connect these judgments to the dimensions of mind perception, we build on \cite{hidalgo_how_2021} by using a regression model predicting the perceived wrongness of a scenario (\textit{W}) as a function of its perceived intention (\textit{I}) and harm (\textit{H}) \cite{hidalgo_how_2021}. 

\begin{equation}
\label{basicmodel}
W = B_1H + B_2I + B_3HI + \eta + \epsilon
\end{equation}

Where $\eta$ represents subject fixed-effects and $\epsilon$ the residual.

This model provides a quantitative representation of moral philosophy. A consequentialist philosophy is one in which outcomes (harm, $H$) dominate the function. This is given by the derivative of wrongness on harm: $dW/dH = B_1 + B_3 I$. A moral philosophy where intention plays a larger role is one in which the role of intention is accentuated. That is, one with a larger $dW/dI = B_2 + B_3 H$. 

\section{Results}\label{sec:results}

\begin{table*}[t]
\renewcommand{\arraystretch}{1.3}
  \caption{Results of moral functions for the human condition and for a combination of all machine conditions}
  \label{sample-table}
  \centering
\begin{adjustbox}{width=1\textwidth}
  \begin{tabular}{lllll}
    \toprule
  \multicolumn{5}{c}{Dependent variable : Wrongness}     \\ 
    \cmidrule(r){1-5}
    \multicolumn{4}{c}{OLS} {felm}  \\ 
    & Human & Machine & Human & Machine \\
    & (1) & (2) &  (1) & (2) \\
    \midrule
    Intention & 0.372*** & 0.216*** & 0.379*** & 0.255***\\
    & (0.055) & (0.035) & (0.056) & (0.034)\\
    Harm & 0.204*** & 0.334*** & 0.177*** & 0.256*** \\
    & (0.045) & (0.021) & (0.047) & (0.020) \\ 
    Intention $\times$ Harm & 0.226*** & 0.158*** & 0.249*** & 0.169*** \\ 
    & (0.068) & (0.043) & (0.070) & (0.041) \\ 
    Constant & 0.161*** & 0.199*** \\ 
    & (0.034) & (0.016) \\ 
    \midrule
    Observations & 1,062 & 4,563 & 1,062 & 4,563 \\ 
    $R^2$ & 0.507 & 0.270 & 0.584 & 0.496 \\ 
    Adjusted $R^2$ & 0.505 & 0.269 & 0.531 & 0.436 \\ 
    Residual Std. Error & 0.247 (df = 1058) & 0.300 (df = 4559) & 0.241 (df = 942) & 0.264 (df = 4071) \\ 
    F Statistic & 362.047*** (df = 3; 1058) & 560.761*** (df = 3; 4559)\\ 
    \bottomrule
    \footnotesize{Note: $^{*}\, p<0.1$; $^{**}\, p<0.05$; $^{***}\, p<0.01$}
    \end {tabular}
    \end{adjustbox}
\end{table*}

Table \ref{sample-table} shows moral functions estimated for the human condition and for a combination of all machine conditions. This confirms the results in \cite{hidalgo_how_2021} showing that the judgment of machine actions follows a relatively consequentialist philosophy compared to the judgment of humans. More precisely, $dW/dH(machines)=0.334+0.158I$ and $dW/dH(humans)=0.204+0.226I$. Thus, $dW/dH(machines)>dW/dH(humans)$ for all values of intention (which ranges from 0 to 1), meaning that people put more weight on a scenario's perceived level of harm $H$ when judging machines. These results hold for all observed ranges of intention using a model with subject fixed effects controlling for all statistic individual characteristics (felm columns in Table 1).
We expand this model by adding perceived agency ($A$) and experience ($E$). Thus, we construct two moral functions of the form:
\begin{equation}
\label{fmfh}
W = f_h (H, I, A, E )\:\: \textrm{and} \:\: W = f_m (H, I, A, E)
\end{equation}

where \textit{h} is the function estimated for the judgment of human actions and \textit{m} is the function estimated for the judgment of machine actions. Going forward, we use $A+$ and $A-$ to denote high agency and low agency for the machine condition and $E+$ and $E-$ to denote high experience and low experience. We expand equation \ref{fmfh} to first-order terms and interactions: 

\begin{multline}   
W = B'_1H + B'_2HA + B'_3HE + B'_4I+ B'_5IA +\\ B'_6IE + B'_7HI+ B'_8HIA + B'_9HIE + \eta + \epsilon
\end{multline}

After grouping terms, we obtain a function equivalent to the one used in \cite{hidalgo_how_2021} but with coefficients that vary depending on the perceived agency ($A$) and experience ($E$).

\begin{multline}
W =(B'_1+ B'_2A + B'_3E) H +(B'_4+ B'_5A + B'_6E) I +\\(B'_7+ B'_8A + B'_9E) HI +  \eta + \epsilon
\end{multline}
\\

For instance, in this model, the coefficient connecting harm ($H$) to wrongness ($W$) is given by: $B_1 = B'_1+ B'_2A + B'_3E$. If increases in agency and experience make people’s judgment of machines more human-like, then we should expect $dW/dI(A+,E+)>dW/dI(A-,E-)$. Also, we can use this model to study if changes in the perceived agency and experience of machines move the function that people use to judge machines closer to the function that people use to judge humans. We can estimate this distance $D$ by calculating the volume between the planes defined by $f_h$ and $f_m$ as:

\begin{equation}
\label{integral}
D(A,E) =\int_0^1 \int_0^1 |f_h(H,I,A,E)-f_m(H,I,A,E)| dH dI
\end{equation}

In sum, if increases in the perceived agency of machines make people's judgment of machines more human-like, then $dW/dI(A+)>dW/dI(A-)$ and $dW/dH(A+)<dW/dH(A-)$. If perceived experience makes people’s judgment of machines more human-like, then $dW/dI(E+)>dW/dI(A-,E-)$ and $dW/dH(E+)<dW/dH(A-)$. 
\begin{table*}[t]
\renewcommand{\arraystretch}{1.3}
  \caption{Regression models considering interactions with agency and experience }
  \label{sample-table2}
  \centering
\begin{adjustbox}{width=0.7\textwidth}
  \begin{tabular}{lll}
    \toprule
  \multicolumn{3}{c}{Dependant variable : Wrongness }     \\ 
    \cmidrule(r){1-3} 
    & OLS & felm  \\ 
    & Machine (1) & Machine (2) \\
    \midrule
    Intention &  0.222*** & 0.445*** \\
    & (0.056) & (0.062) \\
    Harm & 0.380*** & 0.394***  \\
    & (0.024) & (0.034)  \\ 
    Intentional $\times$ Harm & 0.070 & -0.082 \\ 
    & (0.068) & (0.072) \\ 
    Agency $\times$ Intention & -0.120 & -0.317** \\ 
    & (0.107) & (0.132)\\ 
    Agency $\times$ Harm & -0.165*** & -0.270 ***\\ 
     & (0.032) & (0.080)\\ 
    Experience $\times$ Intention & 0.174 & -0.215\\
     & (0.117) & (0.149)\\ 
    Experience $\times$ Harm & 0.089** & -0.127 \\ 
     & (0.041) & (0.097)\\ 
    Agency $\times$ Intention $\times$ Harm & 0.310 ** & 0.403**\\ 
     & (0.132) & (0.157)\\ 
    Experience $\times$ Intention $\times$ Harm & -0.170 & 0.314*\\ 
     & (0.147) & (0.178)\\ 
    Constant & 0.198***  \\ 
    & (0.015)   \\ 
    \midrule
    Observations  & 4,563 & 4,563 \\ 
    $R^2$ & 0.276 & 0.500 \\ 
    Adjusted $R^2$ & 0.274 & 0.493  \\ 
    Residual Std. Error & 0.299 (df = 4553) & 0.263 (df = 4065)  \\ 
    F Statistic & 192.551*** (df = 9; 4553) \\ 
    \bottomrule
    \footnotesize{Note: $^{*}\, p<0.1$; $^{**}\, p<0.05$; $^{***}\, p<0.01$}
    \end {tabular}
    \end{adjustbox}
\end{table*}

Table \ref{sample-table2} presents our main results. Here, we can see the effects of agency and experience on people’s judgment of the four machine conditions. We notice that not all coefficients are significant, so we create reduced models using only significant and robust coefficients (coefficients that are significant in  both the OLS and fixed effect models). Since none of the coefficients involving experience is robust across both specifications, we obtain two equations (using the OLS coefficients):

for $A+$ and $E+$ and for $A+$ and $E-$:
\begin{equation}
W = 0.215H + 0.222I + 0.310HI + 0.198
\end{equation}

and for $A-$ and $E+$ and $A-$ and $E-$:
\begin{equation}
W = 0.38H + 0.222I + 0.198
\end{equation}

As a benchmark, the function of people judging humans (from table 1) is:
\begin{equation}
W=0.204H+0.372I+0.226HI+0.161
\end{equation}

First, we study the effect of perceived agency on the role of harm. That is, we compare $dW/dH(A+)=0.215+0.310I$ with $dW/dH(A-)=0.38$. Setting $dW/dH(A+)=dW/dH(A-)$ we obtain $I=0.53$, meaning that for all values of $I$ less than 0.53 the role of harm on wrongness is smaller for the high agency condition. Since the median value of I is 0.2, this means that--for the majority of our sample--an increase in agency is associated with a reduction of the impact of harm on the judgment of wrongness.
Now we study the effect of perceived agency on the role of intention for each level of perceived experience. That is, we compare $dW/dI(A+)=0.222+0.310H$ with $dW/dI(A-)=0.222$. For the median value of harm, which is $H=0.4$ this gives us $dW/dI(A+)=0.346$ and $dW/dI(A-)=0.222$, meaning that the role of intention increases with agency for the median value of harm. Here, the break-even point is $H=0$, meaning that this is true for all values of $H$.
Overall, these results show that increases in perceived agency are associated with reductions in the role of perceived harm in the judgment of wrongness and increases in the role of perceived intention in the judgment of wrongness.
Next, we estimate the distance between the moral functions describing how people judge machines and humans. We can do this by calculating the integral in equation \ref{integral}. We find that:

\begin{equation}
\label{differences}
    D(A+)=0.0298 < D(A-)=0.0675
\end{equation}

This shows that the distance between the function describing how humans judge high-agency machines and humans is smaller than the distance describing how humans judge low-agency machines and humans. This tells us that machines perceived as having more agency are judged more similar to humans than machines perceived as having lower agency. 
We visualize these moral functions in Figure 3. B. In these plots, the red plane represents people's moral judgment of machines, and the blue plane represents people’s judgment of humans. When agency is low (\textit{A-}), there is a big gap between the blue and red planes for scenarios involving low intention and high harm (severe accidents). This gap is indicative of the aforementioned differences in moral philosophy since it involves judging the severity of an outcome in an accidental situation. This gap is reduced, however, when agency increases. Figure 3. B shows that when the perceived agency of machines is large (\textit{A+}) the red and blue planes become close rto each other at the high harm low intention corner. These observations are confirmed by the results of the integral reported in equation \ref{differences}.
In sum, our findings show that people’s judgment of machines becomes more similar to that of humans when people perceive machines as more capable of exhibiting agency.

\begin{figure*}[htb]
\centering
  \includegraphics[width=0.8\textwidth]{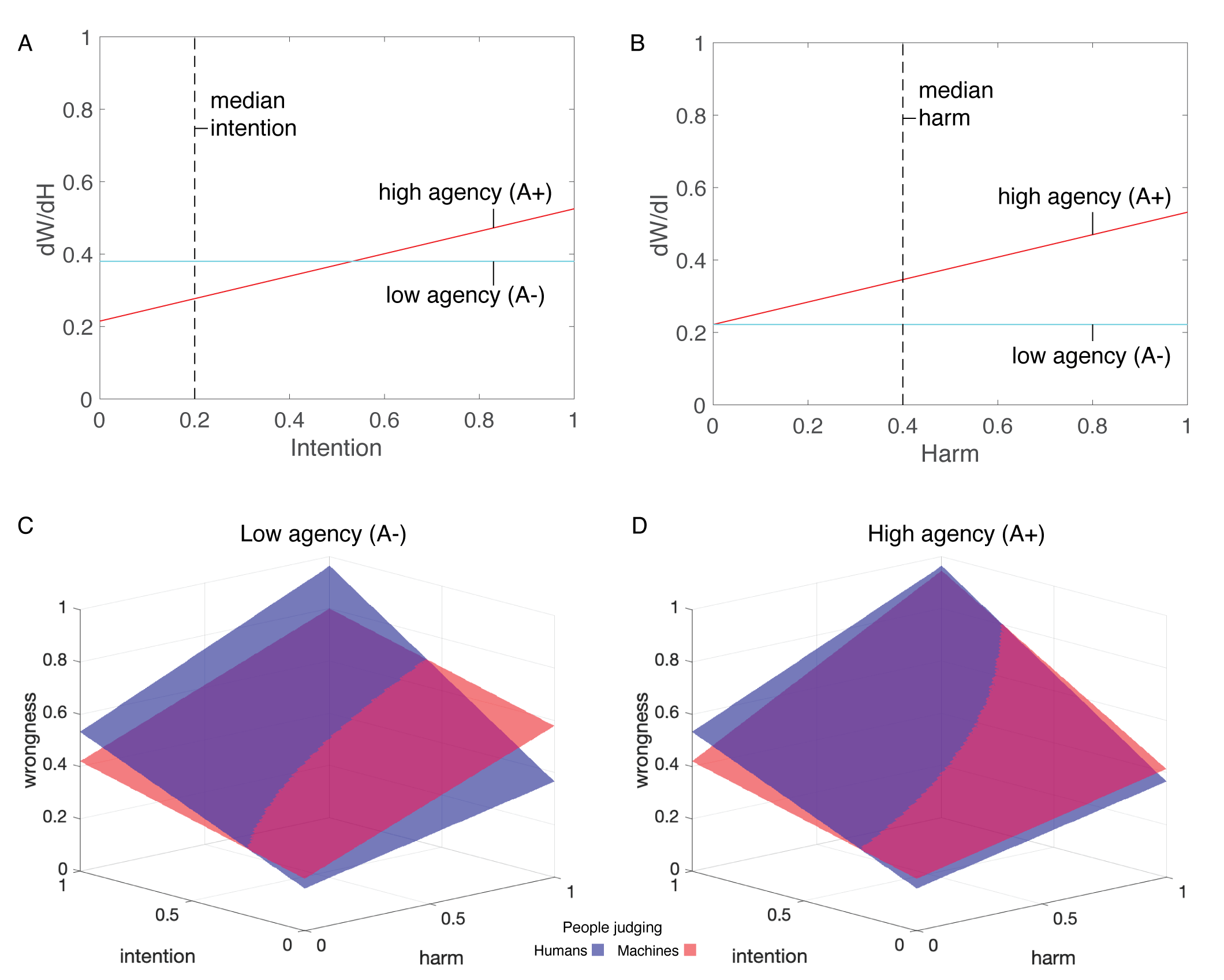}
  \caption{Dependency of wrongness on \textbf{A} harm and \textbf{B} intention. Visualizations of the function’s describing people’s judgments of humans (blue) and machines (red) for \textbf{C} low agency machines and \textbf{D} high agency machines (A+). The blue plane is the same in both charts.}
\label{fig3}
\end{figure*}

\section{Conclusion}\label{sec:discussion}

Researchers have known for some time that people's difference in the judgment of humans and machines goes beyond a simple preference for humans. People judge humans and machines using different moral philosophies \cite{hidalgo_how_2021,malle2015sacrifice}, which can be represented quantitatively using moral functions such as the ones used in this study. But why do people judge humans and machines differently? And can these differences be explained by the way in which people perceive human and machine minds?
In this study, we show that mind perception models of humans and machines can help explain differences in people’s moral judgments. In fact, we find that by manipulating people’s perception of machines to become closer to that of humans we can also bring people’s moral judgment of machines closer to that of humans. When people perceive machines as more human-like, they move from a utilitarian or consequentialist philosophy, where outcomes determine wrongness, to a philosophy where the interaction between intention and harm plays a more predominant role.
We also find that agency seems to matter more than experience when it comes to humanizing people’s judgment of machines. This finding reflects the importance people place on intention when making moral judgments \cite{haidt2018coddling,cushman2008crime,cushman2013development,greene2009pushing,malle1997folk,young2011ignorance}. For instance, Young et al. (2007) found that subjects judged people's actions as equally morally forbidden when they believed their actions would cause harm irrespective of the actual outcome \cite{young2007neural}. But at the same time, this effect could simply reflect the fact that all of the scenarios presented involved machines and humans performing actions. This motivates further research, for instance, exploring whether there may be subcomponents of experience--such as feelings of guilt, regret, and remorse--that contribute to moral reasoning in ways similar to the perceived agency.
But what do these results mean for AI researchers? While controlling these two dimensions (agency and experience) may be difficult, the paper’s contribution is not about how to play with these parameters in practice but about helping us understand how people’s judgment is affected by how people perceive machines, not when they are in direct contact with them, but when they read about their actions. Using these results, an engineering team could run a similar study to test how their technology is perceived, in terms of agency and experience, and with that, anticipate how it would be judged. Similarly, a communications or marketing team could rephrase the way in which a machine is described to steer people's judgment to one mode or the other. This latter example involves more perverse incentives since it could be used to deflect or manipulate public opinion in situations involving the actions of machines.
There are also limitations coming from the sample of participants. The use of a UK sample means that we lack the geographic diversity needed to make claims that generalize to a global audience. Related research studies \cite{awad2018moral,hidalgo_how_2021}, however, give us some reasons to believe that while judgments vary with geographic and demographic characteristics \cite{awad2018moral}, these variations are likely to be second order, meaning that we expect the size effect associated with demographic variables to smaller than the effects observed between scenarios and conditions.
Our study is also limited in that we use only one mind perception model. In principle the differences described in this paper could be explained by other mind perception models, like the warmth competence model we mentioned earlier \cite{cuddy_warmth_2008}. We leave questions, such as whether these alternative mind perception models provide a better explanation, or whether they provide redundant information (because mind perception models are not necessarily independent from each other), to future research. This means our claim is narrow, not only in terms of the demographic representation of participants but on the space of possible mental models.     
Nevertheless, our findings demonstrate the value of modeling humans and machines within the same multidimensional mind perception space to understand the link between perception and moral reasoning. People do not simply judge humans and machines differently. We seem to place machines and humans in different parts of a mind perception space where modes of judgment move together with mind perception. This has significant implications for the design of machines, as it indicates that modifying a person's perception of a machine's agency can affect how that person judges the machine's mistakes. We hope these findings help stimulate further research on the moral philosophy of humans judging machines.
\section*{Acknowledgement}
We acknowledge the support of the Artificial and Natural Intelligence Institute of the University of Toulouse (ANITI) ANR-19-PI3A-0004, the 101086712—LearnData—HORIZON-WIDERA-2022-TALENTS-01 financed by EUROPEAN RESEARCH EXECUTIVE AGENCY (REA) (https://doi.org/10.3030/101086712), IAST funding from the French National Research Agency (ANR) under grant ANR-17-EURE-0010 (Investissements d'Avenir program), the European Lighthouse of AI for Sustainability, HORIZON-CL4-2022-HUMAN-02 project ID: 101120237, and the Science Foundation Ireland and Irish Research Council 21/PATH-A/9615. 

\bibliographystyle{splncs04}
\raggedright
\bibliography{reference}

\end{document}